%% file: Revised_manuscript_without_red.tex
\documentclass[epj]{svjour}
\usepackage{graphicx,epstopdf,multirow,amsmath,rotating,subfigure,isotope,gensymb,amssymb}
\usepackage{comment,commath,enumitem,calc,csquotes,footnote}
\usepackage{graphicx}
\usepackage{dcolumn}
\usepackage{bm}
\usepackage{tabularx}
\usepackage{ulem} 
\usepackage[export]{adjustbox}
\usepackage{float}
\usepackage{braket}
\usepackage{lscape}
\usepackage{lipsum}
\usepackage{longtable}
\graphicspath{{figure/}}
\newcommand\T{\rule{0pt}{3ex}}       
\newcommand\B{\rule[-1.5ex]{0pt}{0pt}} 
\usepackage{xcolor}   

\usepackage{hyperref}
\hypersetup{
    colorlinks=true, 
    linktoc=all,     
    urlcolor  = blue,
    citecolor = blue,
    linkcolor=blue,  
}
\makeatletter
\newcommand{\colorcaption}[2][]{%
  \begingroup%
  \renewcommand{\@caption@fignum@sep}{ (Color online). }%
  \caption[#1]{#2}%
  \endgroup%
}
\makeatother
\begin{document}

\title{Study of S, Cl and Ar isotopes with  $N \geq Z$ using microscopic effective $sd$-shell interactions}

\author{Priyanka Choudhary\thanks{pchoudhary@ph.iitr.ac.in}, Praveen C. Srivastava\thanks{Corresponding author: praveen.srivastava@ph.iitr.ac.in}}

\institute{Department of Physics, Indian Institute of Technology Roorkee, Roorkee-247667, Uttarakhand, India}

\date{Received: date / Revised version: date}

\abstract{
In the present work, newly developed microscopic effective $sd$-valence shell interactions such as chiral next-to-next-to-next-to-leading order (N3LO), $J$-matrix inverse scattering potential (JISP16), Daejeon16 (DJ16), and monopole-modified DJ16 (DJ16A) are employed to study the nuclear structural properties of sulphur, chlorine, and argon isotopes with $N \geq Z$. These interactions are derived using the \textit{ab initio} no-core shell-model and the OLS unitary transformation method. We calculate energy spectra and electromagnetic properties to test the predictive strength of the effective interactions for these heavier $sd$-shell nuclei. For a complete systematic study, we compare the microscopic results with the phenomenological USDB results and experimental data.  By looking at the excitation energies of these nuclei, the DJ16A interaction is found to be {the} most suitable for these $sd$-shell nuclei among all microscopic interactions. The electric quadrupole transition strength and excitation energy of the first $2^+$ state data of even-even sulphur isotopes indicate the presence of the $N=20$ shell closure. Quadrupole moment predictions are also made using these interactions where experimental data are unknown. Magnetic moments are in excellent agreement with the experimental values. The root-mean-square deviations are also calculated to provide an idea of how {accurate} the interactions are.}

\PACS{ 21.60.Cs, 21.30.Fe, 21.10.Dr, 27.30.+t, 21.10.Ky}
\titlerunning{Study of S, Cl and Ar isotopes with  $N \geq Z$ using microscopic effective $sd$-shell interactions}
\authorrunning{P. Choudhary {\it et al.}}
\maketitle

\section{Introduction}
\label{I}
The no-core shell-model (NCSM) is a powerful many-body technique that successfully describes nuclei in the lighter mass region from the first principles \cite{Zheng_1993,Zheng_1994,Navratil_1998,Vary_2004,stetcu_2005,stetcu_2006,Forssen_2008,Navratil_2009,Maris_2009,Barrett_2013,Saxena_2019,Choudhary_2020,Saxena_2020,Choudhary_2023,Choudhary_2023/Jan}. This approach treats all nucleons as active particles interacting via realistic inter-nucleon forces. For nuclei with mass number $A>16$, the NCSM method is unable to provide reliable results due to its vast dimensions of calculations in large basis space, which reach beyond the accessible computation power. Thus, the nuclear properties are described from other approaches using the interaction between nucleons in a restricted valence space. 

The well-known phenomenological $sd$-shell interactions are USD, USDA, and USDB, constructed by fitting two-body matrix elements (TBMEs) from the experimental data \cite{Brown_2006,Richter_2008}. The $sd$-shell  Hamiltonian contains three single-particle {energies of} orbitals $d_{5/2}$, $s_{1/2}$, and $d_{3/2}$ and 63 TBMEs. Two new USD-type interactions are introduced in Ref. \cite{Magilligan_2020}, which are USDC and USDI. The USDC, just like the USDA and USDB, is based on renormalized G-matrix, while the USDI is developed from \textit{ab initio} in-medium similarity renormalization-group (IMSRG) interactions.  For USDA and USDB interactions, the data of the $sd$-shell nuclei with $N\geq Z$ are used in the fit, while for USDC and USDI interactions, the data of $8 \leq Z \leq 20$ and $8 \leq N \leq 20$ nuclei are used which means the proton-rich nuclei are also included. Also, the USDC and USDI interactions include the isospin-breaking terms directly which are necessary to examine isospin-mixing and other isospin symmetry-breaking effects in the $sd$-shell. These interactions can be used to predict the proton-drip line and proton separation energies of the $sd$-shell nuclei. 

Progress has been made in the direction of constructing the effective interactions microscopically. Initially, these effective interactions were obtained from perturbative approaches \cite{KUO196640,HJORTHJENSEN1995125}. Later, \textit{ab initio} methods were used to provide the spectroscopy of nuclei in a non-perturbative manner, wherein a new technique has been proposed in Ref. \cite{Lisetskiy_2008}, where the  interaction is built from the NCSM wave functions followed by the Okubo-Lee-Suzuki (OLS) transformation. For $sd$-shell, based on the above method, developed effective interactions are chiral next-to-next-to-next-to-leading order (N3LO) based on effective field theory \cite{Dikmen_2015}, $J$-matrix inverse scattering potential (JISP16) \cite{Dikmen_2015},  Daejeon16 (DJ16) \cite{Smirnova_2019}, and monopole modified DJ16 (DJ16A) \cite{Smirnova_2019}. Dikmen \textit{et al.} \cite{Dikmen_2015} have performed \textit{ab initio} NCSM calculations using $N_{\mathrm{max}}$ = 4 model space for $A=18$ and 19 nuclei with the JISP16 \cite{Shirokov_2007} and N3LO \cite{Entem_2003} to obtain eigenvalues and eigenvectors of the primary effective Hamiltonian, and then, these eigenvectors are used to solve secondary effective Hamiltonian using second OLS transformation in order to construct the effective $sd$-shell interaction. This Hamiltonian acts in the 0$\hbar$$\Omega$ model space. They found the same low-lying eigenvalues by this effective Hamiltonian as those obtained from the NCSM Hamiltonian in the $N_{\mathrm{max}}$ = 4 space. In Ref. \cite{Smirnova_2019}, authors have implemented DJ16 interaction to study neutron effective single particle energies for oxygen isotopes and $N=14$ isotones, and as per the results, they have carried out some modifications in $T=0$, and $T=1$ centroids of TBMEs of DJ16 \cite{Shirokov_2016} and resulting interaction is coined as DJ16A. Binding energies obtained with DJ16A interaction for O isotopes are in remarkable agreement with the experimental one. {Motivated by the success of these interactions for lower part of the $sd$-shell, we have tested these interactions for heavier $sd$-shell nuclei such as sulphur, chlorine, and argon isotopes.} Earlier, $p$-shell effective interaction is also formed with the INOY potential using this method \cite{Lisetskiy_2008}. 

{There are some other \textit{ab initio} approaches such as IMSRG \cite{Tsukiyama_2011,Bogner_2014,Stroberg_2016,Stroberg_2017} and coupled-cluster \cite{Jansen_2014,Jansen_2016}, developed for the derivation of the microscopic effective interactions in the last decade. In these approaches, two- and three-body interactions are employed. IMSRG and coupled-cluster approaches predict the spectroscopy of nuclei up to $^{208}$Pb \cite{208Pb_nature}.}

The present work aims to perform a systematic shell-model study of sulphur, chlorine, and argon isotopes based on the microscopic effective $sd$-shell interactions. In order to determine the accuracy of these results, a comparison is carried out with phenomenological USDB results. We perform calculations of energy levels, reduced electric quadrupole transition strengths B(E2), quadrupole moments, and magnetic moments.  For all these observables, the root-mean-square (rms) deviations between experimental and theoretical data are also determined.

This paper is organized in the following way: Section \ref{II} briefly introduces the construction of the effective $sd$-shell interaction. We have presented the shell-model results of energy spectra and electromagnetic properties for sulphur, chlorine, and argon chains in Section \ref{III}. Lastly, conclusions are drawn in Section \ref{IV}.

\section{\label{II}Effective two-body $sd$-shell interactions}
The NCSM Hamiltonian for a system of $A$ point-like nucleons, interacting through realistic two-body interaction, is described as
\begin{equation}\label{(1)}
	H_A = \frac{1}{A}\sum_{i<j}^{A}\frac{(\vec{p}_i-\vec{p}_j)^2}{2m} + \sum_{i<j}^{A}V^{\text{NN}}_{ij},
\end{equation}
here, $m$ represents the nucleon mass. The second term of Eq. (\ref{(1)}) contains both nuclear and Coulomb interactions. To obtain eigenvalues and eigenvectors of $H_A$, the Hamiltonian is diagonalized in a many-body spherical harmonic-oscillator (HO) basis. The basis is constrained via a parameter, $N_{\mathrm{max}}$, which represents the maximum number of HO quanta above the minimum configuration of $A$ nucleon system. For the second part of the Eq. \ref{(1)}, if soft \textit{NN} potentials are employed, it is possible to solve the Hamiltonian to get convergent eigenvalues. On the other hand, if one uses standard realistic interactions that generate strong short-range correlations, a large basis space is required to obtain convergent NCSM results. Such large basis spaces are inaccessible computationally, so, one needs to resort to renormalization methods. The OLS transformation \cite{Okubo_1954,Suzuki_1980,Suzuki_1982,Suzuki_1994} is implemented for the present interactions to accelerate the convergence. 

Now, the center-of-mass HO Hamiltonian is added to the starting Hamiltonian and will be subtracted from the final Hamiltonian. Thus, the Hamiltonian will have the form
\begin{equation}\label{(2)}
	\begin{split}
		& \hspace{-1cm}H_{a} + H_{\text{c.m.}}  
		= \sum_{i=1}^{a} \bigg[ \frac{\vec{p}_i^{\,\,2}}{2m}+ \frac{1}{2}m{\Omega}^2 \vec{r}_i^{\,\,2} \bigg] \\
		& \  \hspace{1.5cm}+ \sum_{i<j=1}^{a} \bigg[ V_{ij}^{\text{NN}} - \frac{m {\Omega}^2}{2A} {(\vec{r}_i - \vec{r}_j)}^2 \bigg] . \,
	\end{split}
\end{equation} 
The two-body cluster approximation is used in these NCSM calculations. In the limit $a$ $\rightarrow$$A$, the Hamiltonian (\ref{(2)}) approaches (\ref{(1)}). A primary effective Hamiltonian is constructed by performing the first OLS transformation of the Hamiltonian (\ref{(2)}). These NCSM calculations are solved for $A=18$ system, $^{18}$F, at $N_{\mathrm{max}}$ = 4 and $\hbar$$\Omega$ = 14 MeV \cite{Dikmen_2015} to get eigenvalues and eigenvectors, which are used to generate a second OLS transformation to the valence $sd$-shell. The eigenvalues of the secondary effective Hamiltonian are exactly the same as those obtained from the primary effective Hamiltonian for $^{18}$F. Then, one proceeds to the NCSM calculations for the system of core and core with one nucleon. These calculations are performed for $^{16}$O, $^{17}$F, and $^{17}$O at same parameters as used in the case of $^{18}$F. The calculation with $^{16}$O is performed to determine the core energy. Then, this core energy is subtracted out from the calculations of $^{17}$F and $^{17}$O to obtain effective proton and neutron one-body terms. Subtracting core plus one-body terms from the secondary effective Hamiltonian will give the residual TBMEs for $sd$-valence space. In this way, two-body matrix elements for $sd$-shell interaction are obtained. The detailed procedure for the same is discussed in Refs. \cite{Dikmen_2015,Smirnova_2019}.   

\section{\label{III}Results and discussion}
In this section, we show the results obtained using the shell-model with N3LO, JISP16,  DJ16, and DJ16A effective interactions and compare them with empirical USDB results. The results of energy spectra and electromagnetic properties for the S, Cl, and Ar isotopes starting from the $N=Z$ line and towards more neutron-rich nuclei are presented. The present shell-model calculations are performed using the KSHELL code \cite{Shimizu_2019}. The rms deviations between the experimental and theoretical energies are calculated for all isotopes corresponding to each interaction. For electromagnetic observables, the rms deviations are also determined.

\subsection{Spectroscopy of S isotopes}

\begin{figure*}
	\begin{center}
	\includegraphics[width=8.7cm]{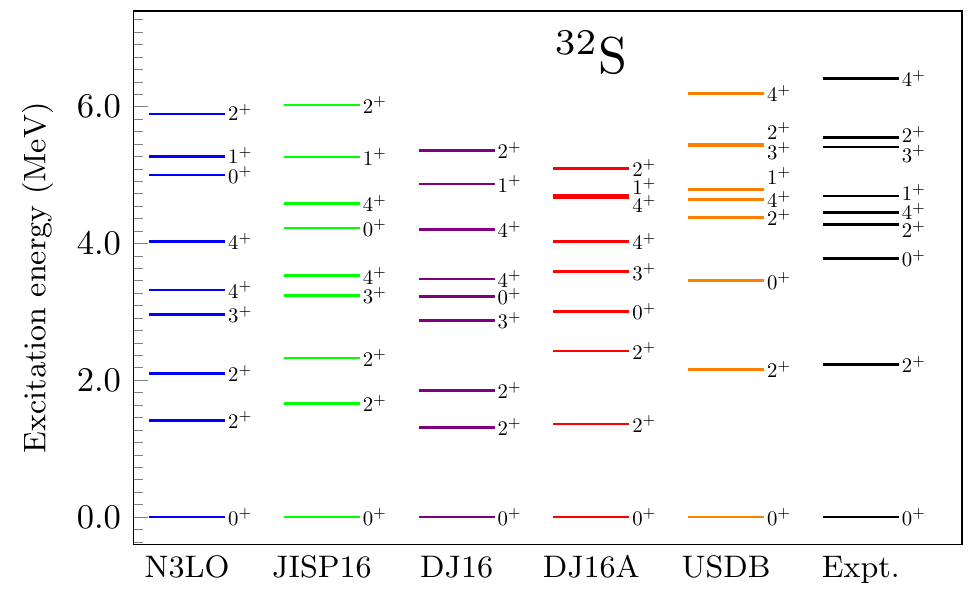}
	\includegraphics[width=8.7cm]{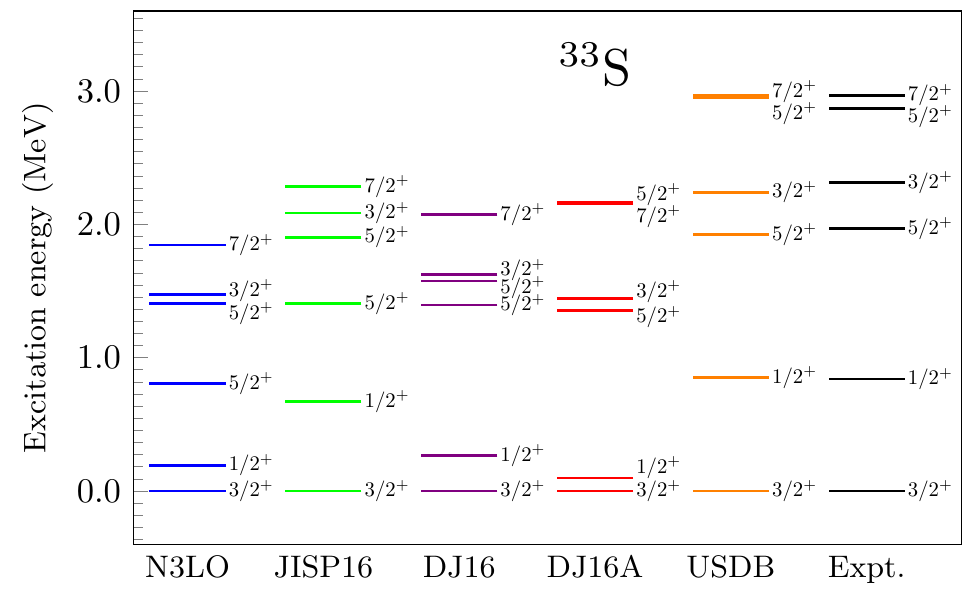}
	\includegraphics[width=8.7cm]{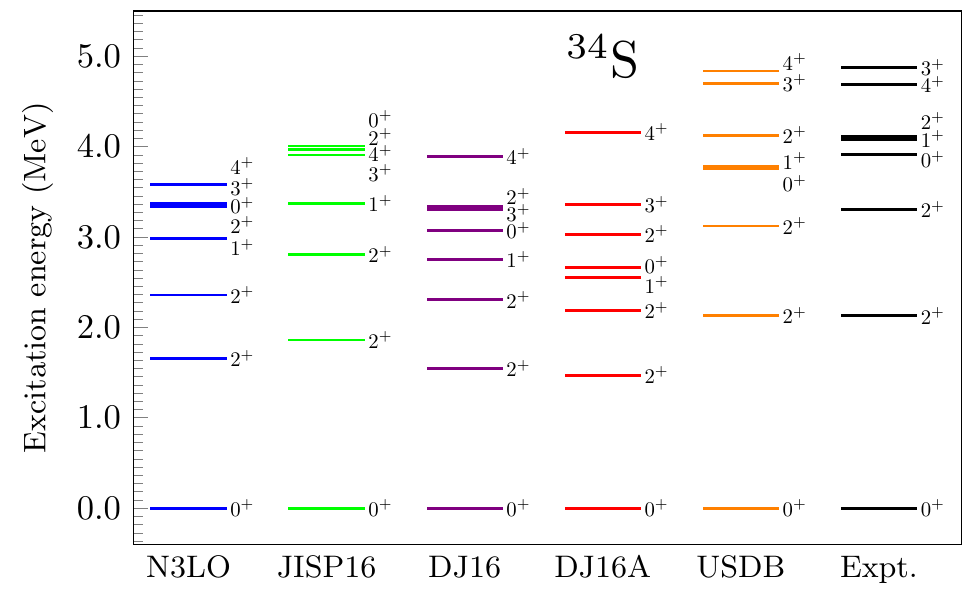}
	\includegraphics[width=8.7cm]{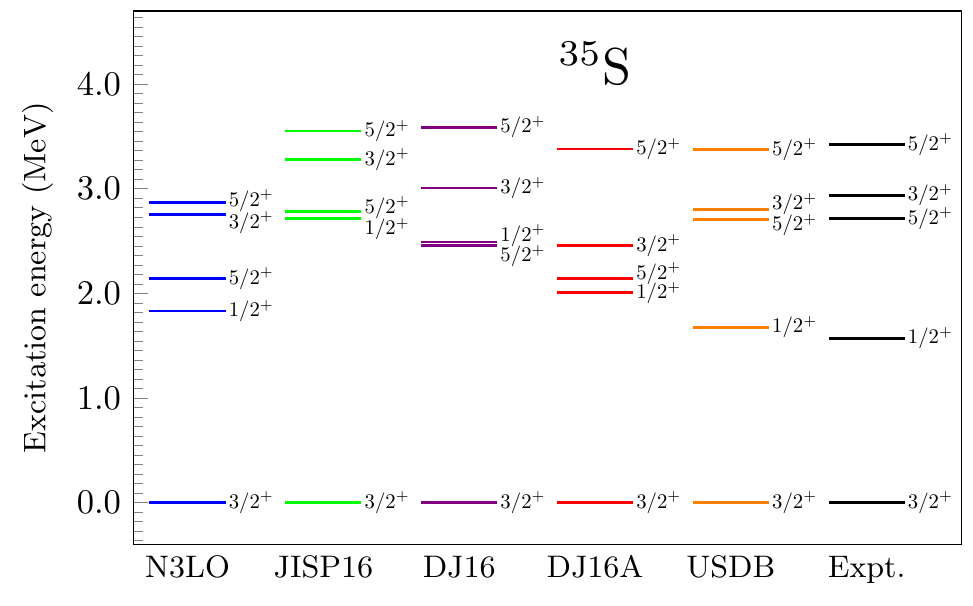}
	\includegraphics[width=8.7cm]{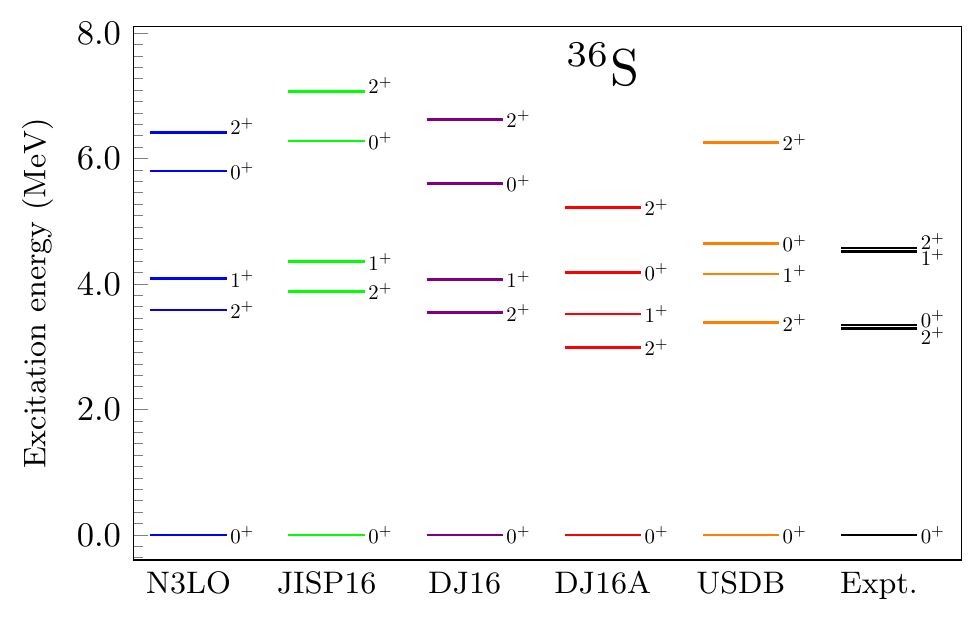}
	\caption{Calculated shell-model low-lying level scheme of sulphur isotopes with $N=16-20$ using N3LO, JISP16, {DJ16,} DJ16A, and USDB interactions in comparison with the experimental data \cite{NNDC}.}
	\label{Fig1}
\end{center}
\end{figure*}

In Fig. \ref{Fig1}, calculated energy levels are reported for $^{32-36}$S isotopes. $^{32}$S has already been studied with these microscopic interactions in Ref. \cite{Smirnova_2019}. Here, we have extended calculations of energy states up to  6.4 MeV.  For $^{33}$S, g.s. is measured as $3/2^{+}$, experimentally, which is well reproduced by the shell-model using all four interactions. The first $1/2^+$ state is predicted at very small excitation energies from N3LO, DJ16, and DJ16A interactions. The low-lying positive parity states obtained from USDB are in the same order as for the experiment. For $^{34}$S, g.s. $0^+$ from the configuration $\ket{\pi (d_{5/2}^6 s_{1/2}^2) \otimes \nu (d_{5/2}^6 s_{1/2}^2 d_{3/2}^2)}$ is obtained with  52.3\% probability using the USDB interaction, while DJ16A interaction predicts this configuration with  40.4\% probability. The lowest $2^+$ state contains the configuration of $\ket{\pi (d_{5/2}^6 s_{1/2}^1 d_{3/2}^1) \otimes \nu (d_{5/2}^6s_{1/2}^2 d_{3/2}^2 )}$ with  32.1\% and  27.0\% probabilities according to the USDB and DJ16A interactions, respectively.  For $^{34}$S, the $d_{3/2}$ proton orbital occupancy increases with excitation energy according to the USDB calculations as 0.67 for $0_1^+$, 0.87 for $2_1^+$, 1.24 for $2_2^+$, 1.65 for $0_2^+$. From DJ16A calculations, the $d_{3/2}$ proton orbital occupancy also increases with excitation energies: 0.91 for $0_1^+$, 1.18 for $2_1^+$, 1.34 for $2_2^+$, 1.70 for $0_2^+$. 
  
\begin{figure*}
  	\begin{center}
  	\includegraphics[width=8.7cm]{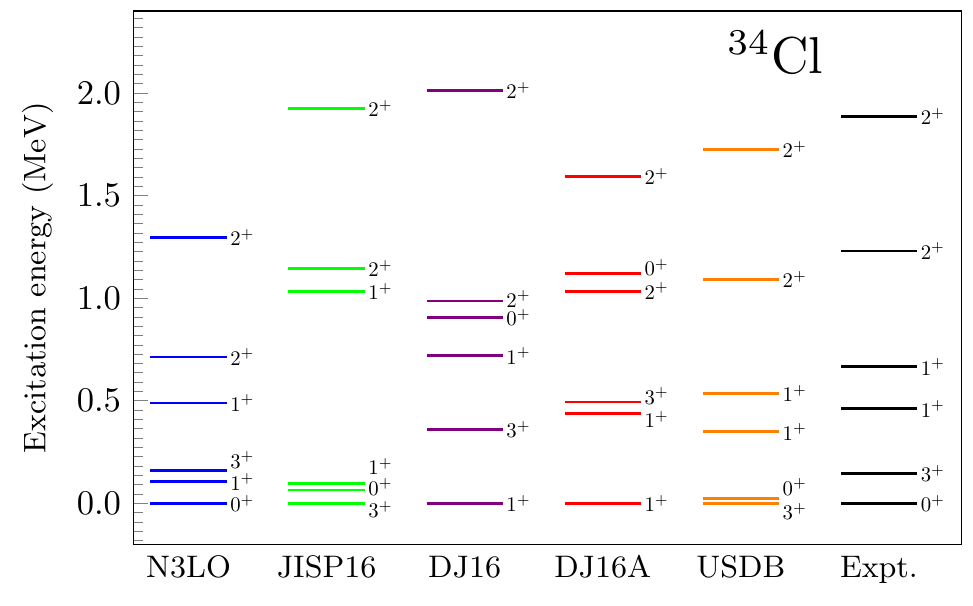}
  	\includegraphics[width=8.7cm]{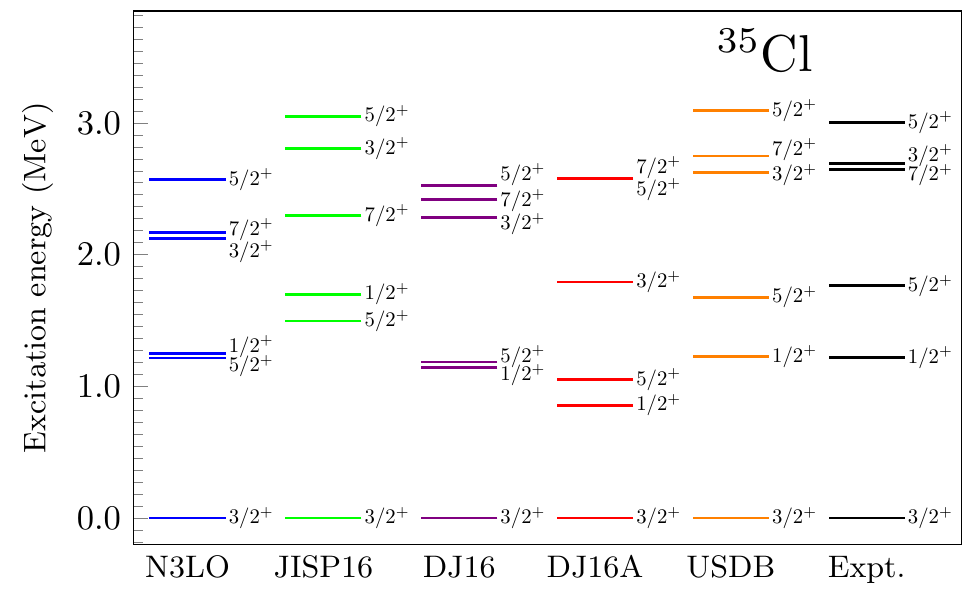}
  	\includegraphics[width=8.7cm]{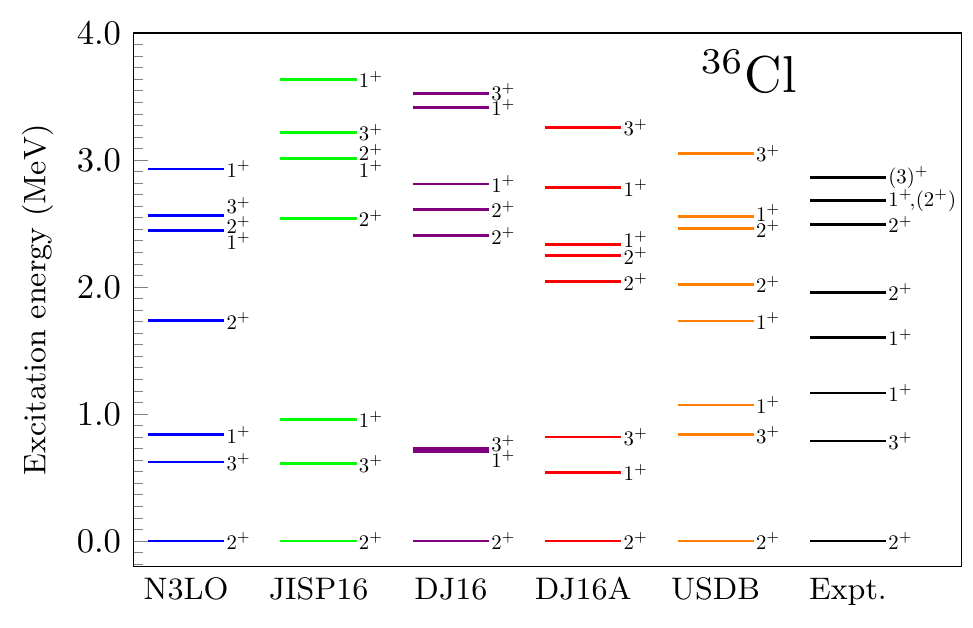}
  	\includegraphics[width=8.7cm]{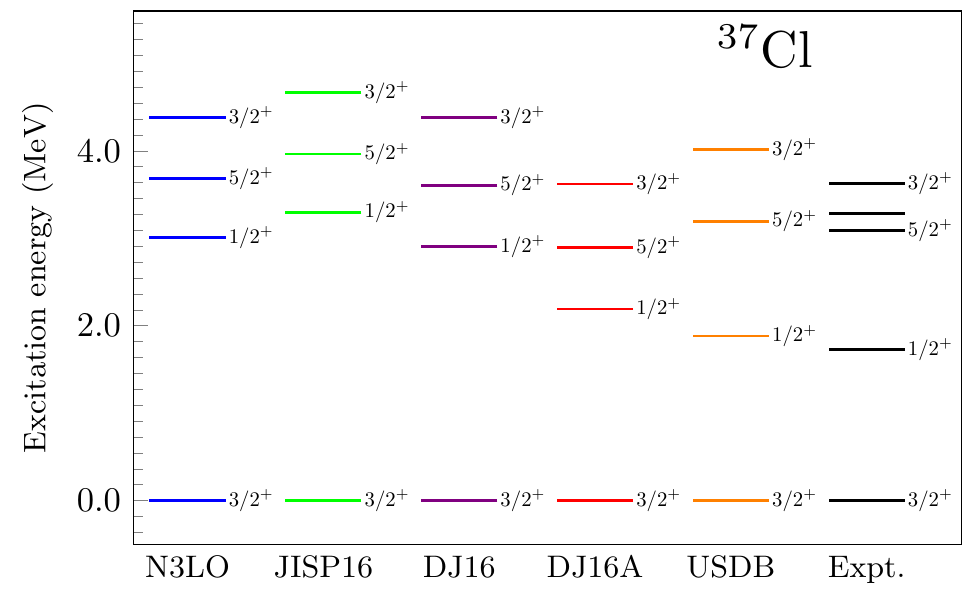}
  	\caption{Calculated shell-model low-lying level scheme of chlorine isotopes with $N=17-20$ using N3LO, JISP16, DJ16A, and USDB interactions in comparison with the experimental data \cite{NNDC}.}
  	\label{Fig2}
  \end{center}
\end{figure*}

Recently in Ref. \cite{Grocutt_2022}, shell-model calculations based on the PSDPF \cite{Bouhelal_2011}, SDPF-U \cite{Nowacki_2009}, and FSU \cite{Lubna_2019} effective interactions have been carried to compare result corresponding to measured lifetime, using the differential recoil-distance method, of the sulphur isotopes with mass numbers $A=35, 36, 37$ and $38$.
For $1/2_1^+$ state, the reported lifetime using PSDPF interaction is about six factor differ from the measured lifetime [3.3(2) $ps$], and it is noted that this discrepancy is due to a problem in the calculation of M1 transition strength \cite{Grocutt_2022}. Our calculated value of lifetime using DJ16A interaction is 0.67 $ps$, while PSDPF interaction gives lifetime as 0.55 $ps$. No experimental mixing ratio is available yet; thus, mixing ratio measurements are needed to give information about the M1 transition rate. { Our calculated value of mixing ratio is 1.10 with DJ16A interaction.} 

Assuming normal filling of shell-model orbitals, one would expect $3/2_{g.s.}^+$, $1/2_1^+$ and $5/2_1^+$ states in $^{35}$S to be due to an unpaired neutron in the $sd$-shell. Direct single-neutron pickup reactions on $^{36}$S have not been found in the literature. From our calculations, we observed that these states with USDB have configurations of \\
$\ket{\pi (d_{5/2}^6  s_{1/2}^2) \otimes \nu (d_{5/2}^6 s_{1/2}^2 d_{3/2}^3)}$,\\ $\ket{\pi (d_{5/2}^6 s_{1/2}^2) \otimes \nu (d_{5/2}^6 s_{1/2}^1 d_{3/2}^4)}$, and\\ $\ket{\pi (d_{5/2}^6 s_{1/2}^1 d_{3/2}^1) \otimes \nu (d_{5/2}^6 s_{1/2}^2 d_{3/2}^3)}$ with  80.6\%,  49.5\%, and  73.4\% probabilities, respectively. The respective probabilities with DJ16A interaction are 72.9\%,  47.6\%, and  62.1\%.  For $1/2_1^+$ state, the other configurations are\\ $\ket{\pi (d_{5/2}^6 s_{1/2}^1 d_{3/2}^1) \otimes \nu (d_{5/2}^6 s_{1/2}^2 d_{3/2}^3)}$ with 19.2\% probability and $\ket{\pi (d_{5/2}^6 d_{3/2}^2) \otimes \nu (d_{5/2}^6 s_{1/2}^1 d_{3/2}^4)}$ with 11.8\% probability corresponding to USDB interaction. While, for DJ16A interaction, the other configurations are\\ $\ket{\pi (d_{5/2}^6 s_{1/2}^1 d_{3/2}^1) \otimes \nu (d_{5/2}^6 s_{1/2}^2 d_{3/2}^3)}$ (14.4\%),\\ $\ket{\pi (d_{5/2}^5 s_{1/2}^2 d_{3/2}^1) \otimes \nu (d_{5/2}^6 s_{1/2}^2 d_{3/2}^3)}$ (8.4\%) and\\ $\ket{\pi (d_{5/2}^6 d_{3/2}^2) \otimes \nu (d_{5/2}^6 s_{1/2}^1 d_{3/2}^4)}$ (7.2\%). Hence, the  $1/2_1^+$ state is not a single-particle state as the g.s. {Shell-model calculations for both positive and negative parity states of $^{35}$S can also be seen in Ref. \cite{Saxena_2017}.}

Now, we focus on $^{36}$S, which has a closed neutron shell $N = 20$. The first $2^{+}$ state is at a high excitation energy of 3.291 MeV. This state has the dominant contribution ($\sim$ 90\%) of $\ket{\pi (d_{5/2}^6 s_{1/2}^1 d_{3/2}^1) \otimes \nu (d_{5/2}^6 s_{1/2}^2 d_{3/2}^4)}$ using the N3LO, JISP16,  DJ16, DJ16A, and USDB interactions. The respective excitation energies of this state are 3.585, 3.878,  3.540, 2.986, and 3.382 MeV. The large excitation energy is due to the significant energy separation between $\pi(s_{1/2})$ and $\pi(d_{3/2})$ orbitals and $N = 20$ shell closure, which leads to the doubly magic character in $^{36}$S. The experimental first yrast $(4^+)$ state at 5.509 MeV in $^{36}$S is predicted around 6-8 MeV from shell-model corresponding to different interactions (not presented in the figure). In Ref. \cite{Grocutt_2022}, the lifetime of yrast $(6^+)$ state is measured for the first time. This state is observed at 6.690 MeV excitation energy, although our shell-model calculations with different interactions give $\sim$ 12-16 MeV, which is too high (not shown in the figure). Two-particle two-hole calculation with FSU interaction is able to reproduce experimental excitation energy and lifetime of this state, which shows the crucial role played by the $pf$-shell orbitals \cite{Grocutt_2022}. The Ref. \cite{Liang_2002} also supports the necessity of inclusion of $pf$-shell above $2_1^+$ state in $^{36}$S.

When we closely inspect the configurations of even-even sulphur isotopes with DJ16A and USDB interactions, we found that the g.s. contains $\ket{(d_{5/2}^6 s_{1/2}^2)}$ proton configuration and neutrons are successively filled into $s_{1/2}$ orbital followed by $d_{3/2}$ as we move from $^{32}$S to $^{36}$S  with maximum probabilities. Also, the first $2^+$ state comes by the excitation of a proton from $s_{1/2}$ to $d_{3/2}$ orbital  for $^{34}$S and $^{36}$S.

\begin{figure*}
 \begin{center}
	\includegraphics[width=8.7cm]{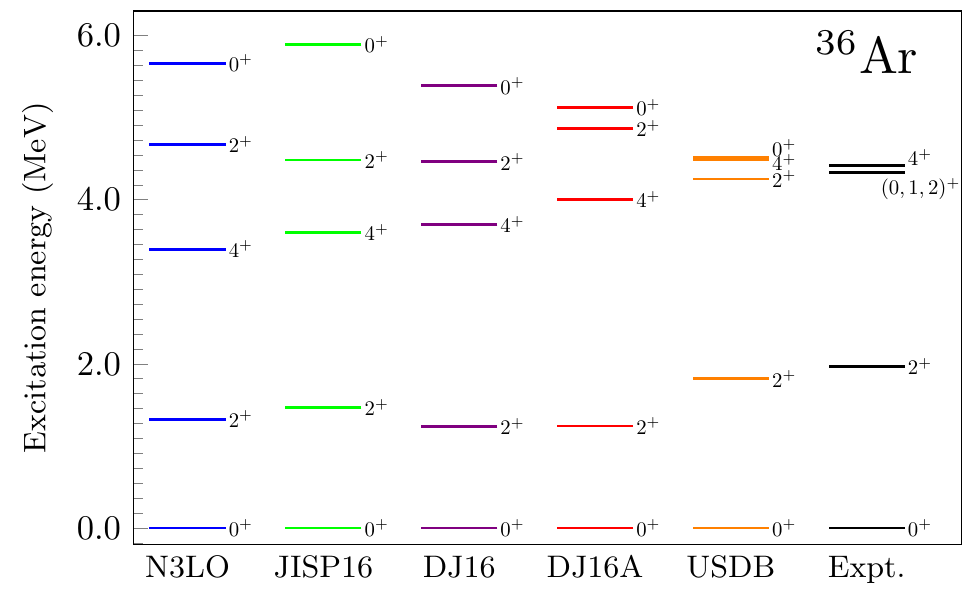}
	\includegraphics[width=8.7cm]{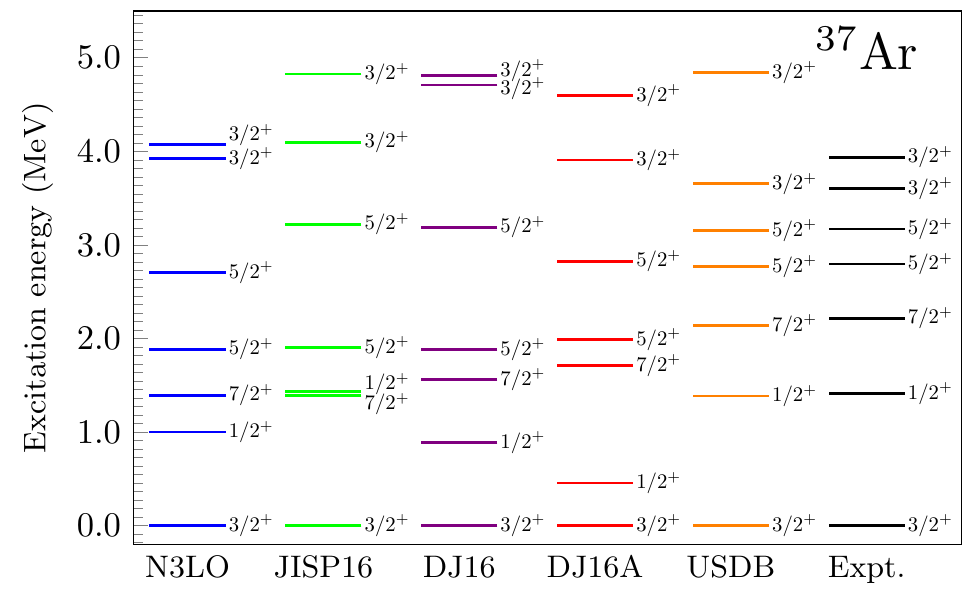}
	\includegraphics[width=8.7cm]{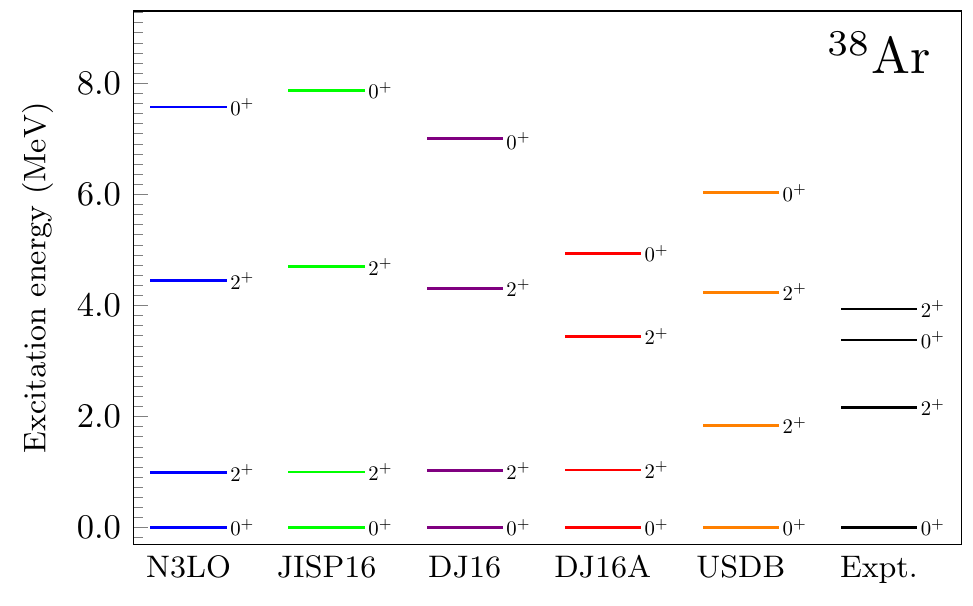}
	\caption{Calculated shell-model low-lying level scheme of argon isotopes with $N=18-20$ using N3LO, JISP16, DJ16A, and USDB interactions in comparison with the experimental data \cite{NNDC}.}
	\label{Fig3}
 \end{center}
\end{figure*}

\subsection{Spectroscopy of Cl isotopes}
\vspace{-0.2cm}
Low-energy spectra for $^{34-37}$Cl are shown in Fig. \ref{Fig2}.
Since $^{34}$Cl is an odd-odd nucleus, it is a difficult case for microscopic interactions. Correct g.s. spin for $^{34}$Cl is only given by the N3LO interaction. The problem may partly be related to J=1, T=0 and J=3, T=0 TBMEs or also in combination with some defects of TBMEs in T=1 channel. The experimental spin-parity of g.s. for  $^{35}$Cl is $3/2^+$, and that of the first excited state is $1/2^+$, which lies 1.219 MeV above the g.s. Looking into the configurations of these calculated states, it seems that the major contribution to the wave function comes from the normal filling of single-particle energy levels. The single particle configurations of $3/2^+$ and $1/2^+$ states are obtained as\\ $\ket{\pi (d_{5/2}^6 s_{1/2}^2 d_{3/2}^1) \otimes \nu (d_{5/2}^6 s_{1/2}^2 d_{3/2}^2)}$ and\\ $\ket{\pi (d_{5/2}^6 s_{1/2}^1 d_{3/2}^2) \otimes \nu (d_{5/2}^6 s_{1/2}^2 d_{3/2}^2)}$, respectively. Ordering of the three lowest states is correctly reproduced by DJ16A interaction for $^{35}$Cl. For $^{36}$Cl, DJ16 and DJ16A interactions give the reverse ordering of states $1_1^+$ and $3_1^+$, while other interactions give the correct ordering. The energy of $3_1^+$ state relative to g.s. is 788 keV, experimentally. Shell-model excitation energies of this state using N3LO, JISP16,  DJ16, DJ16A, and USDB interactions are 622, 608, 726, 818, and 836 keV, respectively. The $^{37}$Cl lies on the boundary of $sd$-shell, the structures of wave functions for $3/2^+$ and $1/2^+$ come from one unpaired proton in $d_{3/2}$ and $s_{1/2}$ orbitals, respectively.  Energies of $1/2^+_1$, $5/2^+_1$, and $3/2^+_2$ states are improved after monopole corrections in DJ16 interaction. In our work, we have calculated only low-lying spin states. High-spin positive and negative states of chlorine nuclei ($A=34, 35, 36, 37$) have been identified and the experimental data have been compared with the shell-model results using USD, PSDPF, sdfp ($s_{1/2}, d_{3/2}, f_{7/2}$ and $p_{3/2}$ orbitals) and SDPFMW interactions in Refs. \cite{Bisoi_2014,Bisoi_2013,Aydin_2012,Ionescu-Bujor_2009}. Obtained results indicate that the $pf$-shell is important for these nuclei.

\subsection{Spectroscopy of Ar isotopes}

We have studied energy spectra of argon isotopes in the range $N = 18 - 20$, which are shown in Fig. \ref{Fig3}. The first excited state $2^+$ lies at 1.970 MeV above the g.s. $0^+$ for $^{36}$Ar. The g.s. is reproduced correctly and excitation energy predicted for $2^+$ using N3LO, JISP16, DJ16, and DJ16A interactions are compressed: 1.326, 1.471, 1.239, and 1.326 MeV, respectively. In $^{36}$Ar, a state at 4.329 MeV could be either $0^+$ or $1^+$ or $2^+$. Thus, shell-model calculations have been performed to determine the spin of this state, and we notice that the $1^+$ state has excitation energy around 6.7-7.0 MeV, so we have not shown this state in the figure. This state is most likely to be $2^+$.

The g.s. of $^{37}$Ar has the configuration of\\ $\ket{\pi (d_{5/2}^6 s_{1/2}^2 d_{3/2}^2) \otimes \nu (d_{5/2}^6 s_{1/2}^2 d_{3/2}^3)}$ with probabilities of\\ 75.1, 78.9, 77.7, 70.7 and 84.5\% according to N3LO, JISP16,  DJ16, DJ16A and USDB interactions, respectively. The JISP16 interaction fails to reproduce the correct order of the experimental $1/2^+$ and $7/2^+$ states. Except for JISP16, all interactions give the same spin-parity of excited states as measured from the experiment. As we go from N3LO to JISP16 to DJ16 to DJ16A interactions, the energy of the first excited $2^+$ state for $^{38}$Ar increases from 0.987 to 1.003 to 1.025 to 1.040 MeV, which is still $\approx$ 1 MeV away from the experimental data (2.167 MeV). Thus, these interactions need to be modified. With WBT interaction \cite{Warburton_1992}, the excitation energy of this state is obtained as 2.032 MeV \cite{Speidel_2006,Speidal_2008}, which agrees well with the experimental value.\\

{\bf The rms deviation:}
\begin{figure}
	\includegraphics[width=\columnwidth]{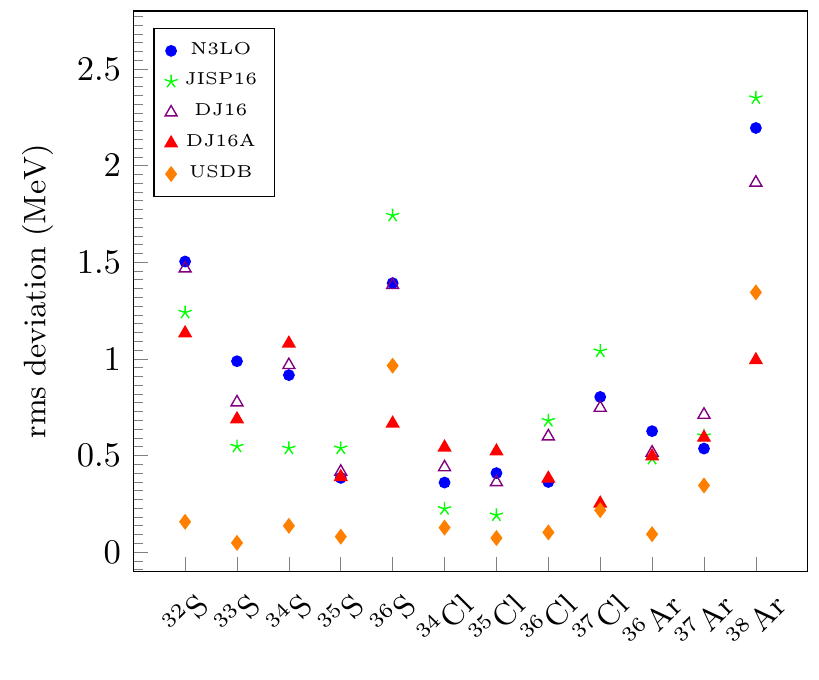}
	\caption{The rms deviations as a function of nuclei for different interactions.}
	\label{Fig4}
\end{figure}
A quantity from the experimental and theoretical values is defined as the rms deviation, which is expressed as follows
\begin{equation}\label{eq}
{\rm rms} = \sqrt{\frac{1}{N}\sum_{i=1}^{N} (E_{\rm exp}^i-E_{\rm th}^i)^2}
\end{equation}
{To quantify the descriptive power} of the microscopic interactions, the rms deviations between experimental and theoretical energies are calculated for all microscopic as well as USDB interactions corresponding to each isotope, which are shown in Fig. \ref{Fig4}. The number of states, which are shown in Figs. \ref{Fig1}-\ref{Fig3} for each isotope, is used to determine the rms deviations. Fig . \ref{Fig4} indicates that sometimes DJ16A results are close to the experimental energies or sometimes energies obtained from JISP16. On the basis of these data, it cannot be said that any one out of these interactions is best suited to describe energies for S, Cl, and Ar isotopes. At $N=20$, there is a large difference between theory and experiment for S and Ar isotopes. For $^{36}$S, the rms deviations between experimental and theoretical results from N3LO, JISP16, DJ16, DJ16A, and USDB are 1.392, 1.742, 1.382, 0.666, and 0.965, respectively. For $^{35}$Cl, the rms deviation is less than 0.5 MeV from all interactions. The largest rms deviation is obtained for $^{38}$Ar. The rms deviation between experimental and theoretical data is 437 keV for USDB interaction with 72 energy levels of 12 nuclei having $N\geq Z$. Using microscopic interactions N3LO, JISP16, DJ16, and DJ16A, the rms deviations are 997, 976, 966, and 735 keV, respectively. Hence, we can say that the DJ16A interaction is the most suitable interaction among all microscopic interactions for excitation energies. Monopole-modifications in the DJ16 interaction reduce the deviation between the experiment and theory.

Two-nucleon interaction, inspired by the effective field theory (EFT), N3LO, is inadequate to describe the spectroscopy of $sd$-shell nuclei due to omitted three-nucleon (3$N$) forces. The DJ16 and JISP16 interactions are constructed in such a way that these can be used without the inclusion of 3$N$ forces but are fitted only with the data of $A\leq 16$ nuclei. This leads to inaccurate results of $sd$-shell nuclei with $A> 16$, which indicates the need for tuning of TBMEs. Minimal monopole modification had been done by Smirnova \textit{et al.} in DJ16 interaction, this improves the energy spectra of heavier $sd$-shell nuclei. Still, this is insufficient to fully explain the energy spectra. Further, some non-monopole components such as pairing, quadrupole-quadrupole correlations are required to be modified. Also, the discrepancies between theory and experiments could be diminished by incorporating the 3$N$ forces in the NCSM calculations and/or extending the model space in the construction of the primary effective Hamiltonian.

\begin{table*}
	\renewcommand{\arraystretch}{1.0}\tabcolsep 0.3cm
	\caption{Calculated shell-model reduced electric quadrupole transition strengths of sulphur, chlorine and argon isotopes  using different interactions with {standard effective} charges $e_p=1.5e$,  $e_n=0.5e$ and optimized effective charges $e_p=1.36e$,  $e_n=0.45e$ \cite{Richter_2008}. These values are separated by `/'. Experimental  B(E2) values are taken from  \cite{NNDC}. The B(E2) values are in $e^2$fm$^4$.}
	\label{tab1}
	\begin{center}
		\begin{tabular}{ccccccccc}
			\hline
			\hline
			Nuclei & A	&	$J_i^{\pi} \rightarrow J_f^{\pi}$	&	N3LO	&	JISP16	& DJ16 &	DJ16A		&	USDB&	Exp.	\T\B \\
			\hline
		S	&	32	&	$2_{1}^{+}$$\rightarrow$$0_{1}^{+}$ & 117.3/96.1  & 107.7/88.2  & 107.7/88.2& 84.4/69.1 &  60.1/49.2 & 59.6(12) \cite{Speidel_2006} \T\B \\
			&	&	$4_{1}^{+}$$\rightarrow$$2_{1}^{+}$ & 121.6/99.6  & 109.0/89.3  & 136.5/111.8 & 121.2/99.3     & 85.5/70.1 &  84(18)  \T\B \\
			&		&	$0_{2}^{+}$$\rightarrow$$2_{1}^{+}$  & 27.0/22.1    & 57.2/46.9 & 32.6/26.7  & 19.0/15.6 & 67.0/54.8 &  71.2(72)  \T\B \\	
			&	33	&	$1/2_{1}^{+}$$\rightarrow$$3/2_{1}^{+}$  & 55.6/45.5   & 41.7/34.2 & 15.0/12.3 & 62.7/51.4    &  24.0/19.7 &  38(25)\T\B \\
			&		&	$5/2_{1}^{+}$$\rightarrow$$3/2_{1}^{+}$ & 13.1/10.8   & 5.2/4.2 & 59.7/48.9 &  61.2/50.1    & 61.8/50.5 &  50(19)\T\B \\
			&	34	&	$2_{1}^{+}$$\rightarrow$$0_{1}^{+}$& 73.2/60.0      & 60.4/49.4  & 58.6/48.0 & 60.0/49.2   & 45.9/37.6 &  40.8(10) \T\B \\
			&		&	$0_{2}^{+}$$\rightarrow$$2_{1}^{+}$ & 25.1/20.6 & 25.9/21.3 & 26.2/21.6 & 19.2/15.8  & 26.3/21.6 &  27.5(46)    \T\B \\	
			&	35	&	$1/2_{1}^{+}$$\rightarrow$$3/2_{1}^{+}$  & 62.1/50.9  & 57.7/47.3 & 54.0/44.2& 47.7/39.1    & 39.7/32.5 &  25.8(48)\T\B \\
			&		&	$5/2_{1}^{+}$$\rightarrow$$3/2_{1}^{+}$ & 24.0/19.7  & 24.9/20.5 & 20.1/16.5 & 24.7/20.3     &  32.6/26.7 &  48(20)\T\B \\
			&	36	&	$2_{1}^{+}$$\rightarrow$$0_{1}^{+}$ & 21.5/17.6   & 22.6/18.6  & 21.1/17.4 & 23.3/19.2    & 26.2/21.6 &  17.6(15) \cite{Grocutt_2022} \T\B \\		
		Cl	&	35	&	$1/2_{1}^{+}$$\rightarrow$$3/2_{1}^{+}$  & 8.4/6.9  & 14.1/11.6  & 13.8/11.4& 9.5/7.8    &  13.8/11.3 &  18.4(34)\T\B \\
			&	&	$5/2_{1}^{+}$$\rightarrow$$3/2_{1}^{+}$ & 101.2/82.9  & 89.9/73.6 & 97.0/79.5 & 109.0/89.3    &  82.3/67.4 &  76.2(82) \T\B \\
			&	&	$7/2_{1}^{+}$$\rightarrow$$3/2_{1}^{+}$ & 36.8/30.1   & 33.1/27.1 & 35.1/28.7 & 36.7/30.0    &  24.6/20.1 &  25.2(34)\T\B \\
	
            &	36	&	$3_{1}^{+}$$\rightarrow$$2_{1}^{+}$ & 44.8/36.7     & 43.1/35.4 & 47.9/39.2 & 49.7/40.8  &  35.7/29.2 &  71(28) \T\B \\
			&		&	$1_{1}^{+}$$\rightarrow$$2_{1}^{+}$ & 58.8/48.2     & 57.2/46.9 & 56.5/46.4 & 57.6/47.2  & 52.6/43.2  &  3.5(13)\T\B \\
				&	37	&	$1/2_{1}^{+}$$\rightarrow$$3/2_{1}^{+}$ & 19.8/16.3  & 21.4/17.6 &22.6/18.6 & 23.4/19.2     &  20.6/16.9 &  16.8(37)\T\B \\
			&		&	$5/2_{1}^{+}$$\rightarrow$$3/2_{1}^{+}$ & 38.9/32.0  & 40.6/33.4 & 39.3/32.3 & 42.9/35.3     &  44.4/36.5 &  19.0(44)\T\B \\	
		Ar	&	36	&	$2_{1}^{+}$$\rightarrow$$0_{1}^{+}$& 71.8/58.8     & 67.2/55.1 & 70.6/57.8 & 76.5/62.7     &  64.8/53.1 &  57.9(35)\T\B \\
			&		&	$4_{1}^{+}$$\rightarrow$$2_{1}^{+}$& 86.6/70.9     & 80.3/65.8 & 86.3/70.7 & 93.5/76.6    & 80.7/66.1  &  85(11)\T\B \\
			&	37	&	$7/2_{1}^{+}$$\rightarrow$$3/2_{1}^{+}$& 41.0/33.6   & 36.9/30.3 & 40.4/33.1 & 45.9/37.7     &  38.5/31.6 &  33.7(37)\T\B \\
			&	38	&	$2_{1}^{+}$$\rightarrow$$0_{1}^{+}$& 30.0/24.7     & 29.4/24.2 & 29.0/23.8 & 30.1/24.8   &  31.0/25.5 &  25.8(12)\T\B \\
			\noalign{\smallskip}\hline\hline\noalign{\smallskip}
		\end{tabular}
\end{center}
\end{table*}

\begin{figure*}
	\begin{center}
		\includegraphics[width=\columnwidth]{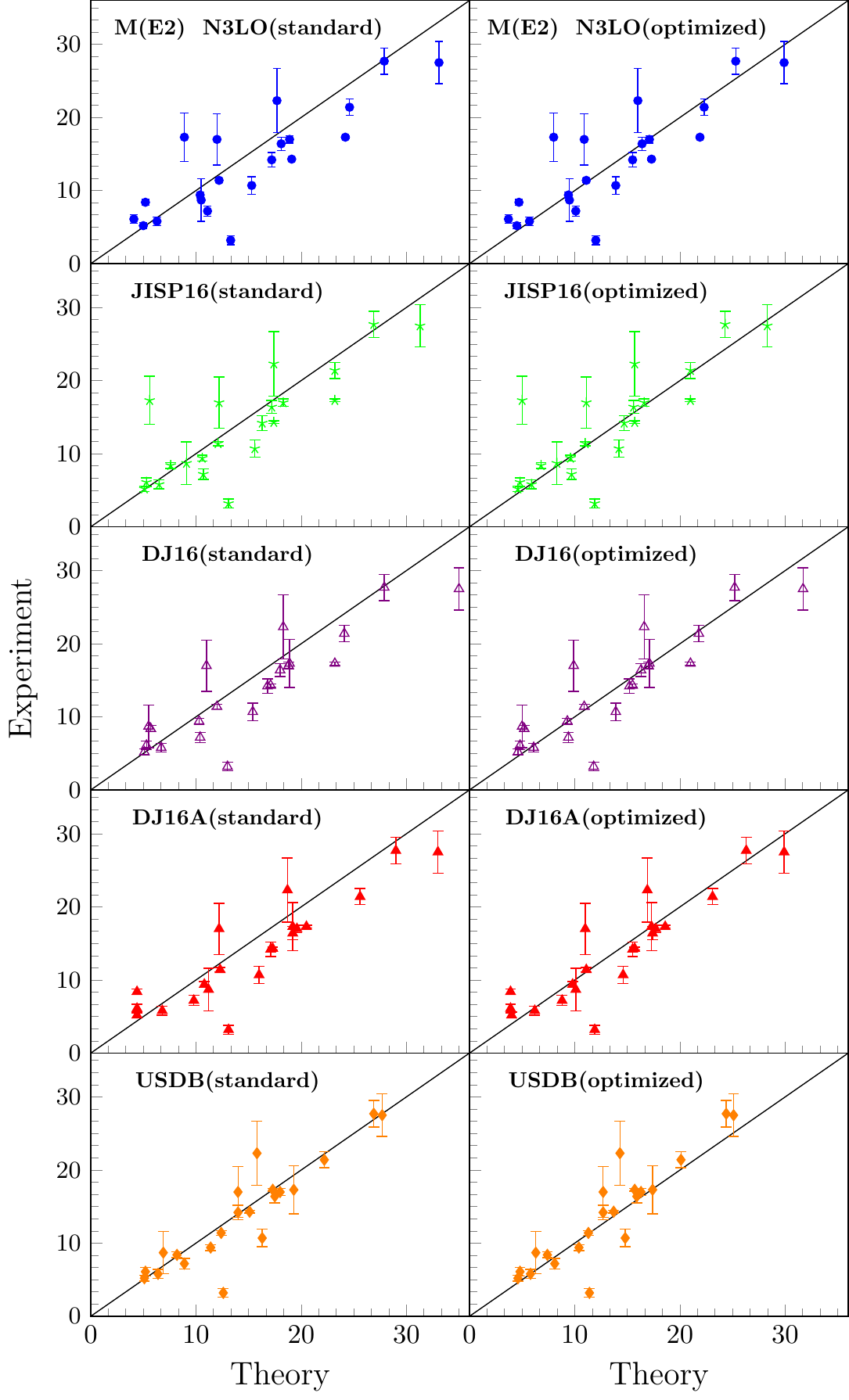}
		\caption{{Comparison of absolute values of experimental and theoretical reduced E2 matrix elements with standard and optimized effective values for the N3LO, JISP16, DJ16, DJ16A, and USDB interactions.}}
		\label{Fig5}
	\end{center}
\end{figure*}

\subsection{Electromagnetic properties}

We have now examined the microscopic interactions by determining the electromagnetic observables of the S, Cl, and Ar isotopes. The B(E2) transition strength obtained with microscopic as well as USDB interactions are shown in Table \ref{tab1}.  We have performed two sets of calculations: one with {standard values of effective} charges $e_p=1.5e$,  $e_n=0.5e$ and the other with optimized effective charges $e_p=1.36e$,  $e_n=0.45e$ \cite{Richter_2008}. The results corresponding to these two sets are reported in Table \ref{tab1}, separated by `/'. The microscopic interactions predict the most collective nuclei as $^{32}$S, which agrees with the experiment. Theoretically obtained B(E2; $2^+$ $\rightarrow$ $0^+$) decreases with addition of neutrons to the $^{32}$S up to $N=20$, which supports the experimental trend. The transition from $1/2_1^+$ to $3/2_1^+$ for $^{35}$S is assigned as M1 character \cite{Aydin_2014}, although in Refs. \cite{Go_2021,Grocutt_2022}, it was known that this is mixed M1-E2 transition. Here, we have reported E2 decay strength for this transition using microscopic interactions, shown in Table \ref{tab1}, and found a very strong transition strength compared to the experimental one. When we move to the $N=20$ shell closure, observed B(E2) value from first $2^+$ to g.s. suddenly drops in the S chain. The effect of the $N=20$ shell gap is clearly seen from the B(E2) values and also from the excitation energy of $2^+$. In the case of  \textsuperscript{36}S, the experimental value is taken from Ref. \cite{Grocutt_2022}. A good agreement between theory and experiment has been achieved for the B(E2) value. Optimized effective charges decrease the B(E2) transition strengths, hence, better results are obtained except for some transitions. These transitions are $0_{2}^{+}$$\rightarrow$$2_{1}^{+}$ for $^{32}$S, $0_{2}^{+}$$\rightarrow$$2_{1}^{+}$ for $^{34}$S, $5/2_{1}^{+}$$\rightarrow$$3/2_{1}^{+}$ for $^{35}$S, $1/2_{1}^{+}$$\rightarrow$$3/2_{1}^{+}$ for $^{35}$Cl, 	$3_{1}^{+}$$\rightarrow$$2_{1}^{+}$ for $^{36}$Cl, and 	$4_{1}^{+}$$\rightarrow$$2_{1}^{+}$ for $^{36}$Ar.  For $^{36}$Cl, weak transition strength from $3_1^+$ to $2_1^+$ is found with all interactions, although strong transition strength is obtained for $1_1^+$ to $2_1^+$ transition compared to the experiment. In the case of $^{37}$Cl, the B(E2) values calculated using different interactions for transition $1/2_1^+$$\rightarrow$$3/2_1^+$ are in reasonable agreement with the experimental value.  At $N=20$, E2 transition strength is decreased with all interactions in the Ar chain, which are in good agreement with the experimental data. Also, we obtain enhancement in the collectivity in the transition strength for some Ar isotopes than the experimental one. A plot of absolute values of experimental reduced E2 matrix elements extracted from the B(E2) data versus theoretical one is presented in Fig. \ref{Fig5}.

\begin{table*}
	\renewcommand{\arraystretch}{1.0}\tabcolsep 0.2cm
	\centering
	\caption{Calculated shell-model electric quadrupole moments of sulphur, chlorine and argon isotopes with {standard effective} charges $e_p=1.5e$,  $e_n=0.5e$ and optimized effective charges $e_p=1.36e$,  $e_n=0.45e$ \cite{Richter_2008}. These values are separated by `/'. Experimental moments are taken from  \cite{Stone_2016,IAEA}.} 
	\begin{tabular}{lcccccccc}
		\hline
		\hline
		&                     &     & \multicolumn{5}{c}{Quadrupole Moment (eb)}   \T\B\\
		\cline{4-9}
		Nuclei  & $A$  & J$^{\pi}$ & N3LO& JISP16 & DJ16 & DJ16A & USDB 	& Exp.\T\B\\
		\hline
		S&	32	&	2$^+$ & 0.097/0.088   & 0.055/0.050    & $-$0.009/$-$0.008 & $-$0.138/$-$0.125 & $$-$$0.128/$-$0.116  & $-$0.16(2) \T\B\\
		&  & 4$^+$ & 0.043/0.039 & 0.033/0.030& 0.038/0.034 & $-$0.051/$-$0.046 &$-$0.060/$-$0.054 & NA\T\B\\		
		&	33	&  $3/2^{+}$   & $-$0.134/$-$0.121  & $-$0.115/$-$0.104 & $-$0.123/$-$0.111 & $-$0.121/$-$0.110  & $-$0.073/$-$0.066     & $-$0.0678(13) \T\B\\	
		&	34	&	$2^{+}$  & 0.039/0.035    & 0.042/0.038  & $-$0.034/$-$0.031 & $-$0.059/$-$0.053 & 0.044/0.040   & 0.04(3)  \T\B\\
		&	35	&	$3/2^{+}$   & 0.078/0.070   & 0.074/0.067  & 0.078/0.070 & 0.080/0.072  & 0.057/0.051   & 0.0471(9) \T\B\\
		&	36	& 	$2^{+}$   & $-$0.106/$-$0.096   & $-$0.102/$-$0.092  & $-$0.099/$-$0.090 & $-$0.103/$-$0.094  & $-$0.103/$-$0.094     & NA
		\T\B\\							
		Cl	&	35		& $3/2^+$	&	$-$0.105/$-$0.095	&	$-$0.107/$-$0.097	& $-$0.112/$-$0.101 &	$-$0.108/$-$0.097	&	$-$0.097/$-$0.088						&	$-$0.0817(8)	\T\B\\	
		&	36		& $2^+$		&	$-$0.019/$-$0.017	&	$-$0.020/$-$0.018	&	$-$0.048/$-$0.043 & $-$0.068/$-$0.061	&	$-$0.017/$-$0.015	&	$-$0.178(4)	\T\B\\
		&	37		& $3/2^+$		&	$-$0.089/$-$0.081	&	$-$0.089/$-$0.081	&	$-$0.089/$-$0.080 & $-$0.089/$-$0.081	&	$-$0.086/$-$0.078&	$-$0.0644(6)	\T\B\\			
		Ar	&	36		& $2^+$		&	0.167/0.152	&	0.152/0.138	& 0.167/0.151 & 	0.178/0.161	&	0.152/0.137	&	0.11(6)	\T\B\\			
		&	37		& $3/2^+$	&	0.097/0.088	&	0.095/0.086	& 0.097/0.088 &	0.103/0.093	&	0.090/0.081 	&	0.076(9)	\T\B\\			
		&	38		& $2_1^+$	&	0.011/0.010 &	0.009/0.008	&	0.007/0.006 & 0.014/0.012	&	0.029/0.026	&	NA	\T\B\\									
		&	& $2_2^+$	&	0.078/0.071	&	0.078/0.081	& 0.079/0.072 &	0.071/0.064	&	0.059/0.053	&	NA \T\B\\												
		\noalign{\smallskip}\hline\hline\noalign{\smallskip}
	\end{tabular}	
	\label{tab2}
\end{table*}

\begin{table*}
	\renewcommand{\arraystretch}{1.0}\tabcolsep 0.3cm
	\centering
	\caption{Calculated shell-model magnetic dipole moments of sulphur, chlorine and argon isotopes with {g$^{\rm free}_{l,s}$ (free nucleon g-factors)} and g$_{l,s}^{p,\rm eff}$ = (1.159, 5.15), g$_{l,s}^{n,\rm eff}$ = ($-$0.09, $-$3.55) \cite{Richter_2008}. These two sets are separated by `/'. Experimental moments are taken from  \cite{Stone_2016,IAEA}.} 	
	\begin{tabular}{lcccccccc}
		\hline\hline
		&                     &       & \multicolumn{5}{c}{Magnetic Moment ($\mu_N$)} \T\B\\
		\cline{4-9}
		Nuclei  & $A$  & J$^{\pi}$ & N3LO& JISP16 & DJ16 &  DJ16A & USDB 	& Exp. \T\B\\
		\hline
		S&	32	&	2$^+$ 
		& 1.011/1.077 & 1.014/1.079   &  1.015/1.080 & 1.015/1.080 & 1.006/1.073 & 0.9(2) \T\B\\
		&  & 4$^+$ 	& 2.026/2.156 & 2.031/2.160& 2.037/2.164 &2.041/2.167 & 2.029/2.158 & 1.6(6)\T\B\\		
		&	33	&  $3/2^{+}$  & 0.248/0.263   & 0.390/0.349 & 0.364/0.359 & 0.542/0.526    & 0.792/0.653 	& 0.6438212(14) \T\B\\	
		&	34	&	$2^{+}$  & 1.062/1.152  & 1.029/1.067  & 1.185/1.308 &  1.297/1.432   & 1.093/1.067 & 1.0(2) \T\B\\
		&	35	&	$3/2^{+}$  & 1.163/1.012   & 1.093/0.938 & 1.026/0.884 & 1.015/0.881    & 1.112/0.923 & 1.00(4) \T\B\\
		&	36	& 	$2^{+}$  & 2.482/2.738    & 2.542/2.790 & 2.482/2.737 & 2.516/2.767    & 2.381/2.650 & 2.6(10) \cite{Speidal_2008}\T\B\\							
		Cl	&	35		& $3/2^+$	&	0.836/1.061	&	0.855/1.081	& 0.965/1.170 &	0.880/1.092	&	0.589/0.861&	0.8218743(4)	\T\B\\	
		&	36		& $2^+$		&	1.553/1.646	&	1.497/1.588	& 1.306/1.415	&1.167/1.303&	1.173/1.287 &	1.28547(5)	\T\B\\
		&	37		& $3/2^+$	&	0.610/0.964	&	0.606/0.960	&0.607/0.962	&0.588/0.944	&	0.352/0.739&	0.6841236(4)	\T\B\\						
		Ar	&	36		& $2^+$		&	0.998/1.068	&	0.989/1.061	& 0.991/1.063	&0.996/1.067	&	0.973/1.050 &	1.0(4)	\T\B\\			
		&	37		& $3/2^+$	&	1.472/1.350	&	1.436/1.307	&1.392/1.281	&1.446/1.355	&	1.225/1.094 &	1.145(5)	\T\B\\			
		&	38		& $2_1^+$   &	0.626/1.122	&	0.605/1.104	& 0.586/1.087	&0.588/1.089	&	0.462/0.979   &	0.5(2)	\T\B\\									
		&	& $2_2^+$	&	2.525/2.775&	2.573/2.817	& 2.578/2.821	&2.585/2.827		&	2.552/2.798&	2.2(22)	\T\B\\															
		\noalign{\smallskip}\hline\hline\noalign{\smallskip}
	\end{tabular}	
	\label{tab3}
\end{table*}

\begin{figure*}
	\begin{center}
		\includegraphics[width=\columnwidth]{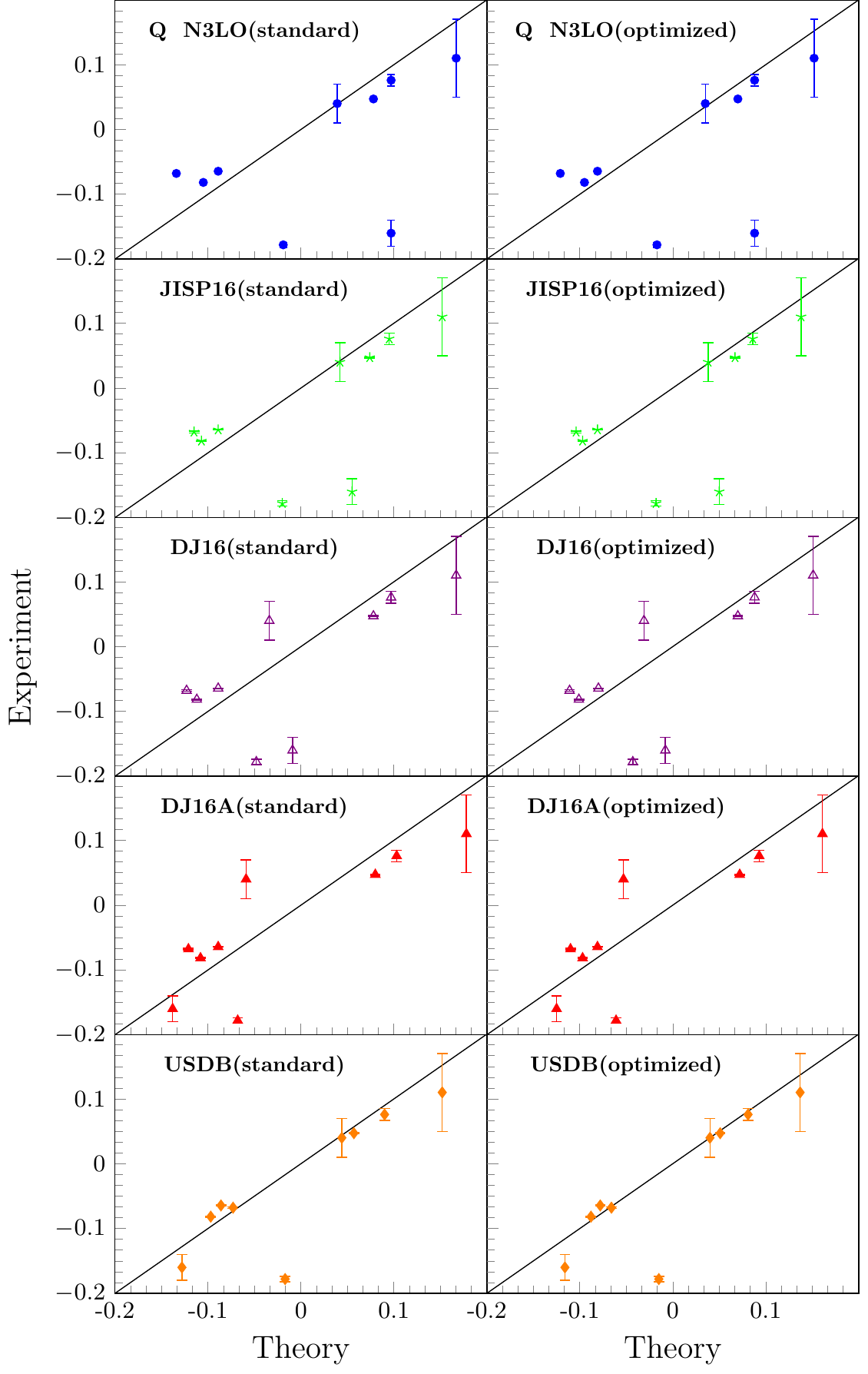}
		\caption{{Comparison of experimental and theoretical quadrupole moments with standard and optimized effective values for the N3LO, JISP16, DJ16, DJ16A, and USDB interactions.}}
		\label{Fig6}
	\end{center}
\end{figure*}

\begin{figure*}
	\begin{center}
		\includegraphics[width=\columnwidth]{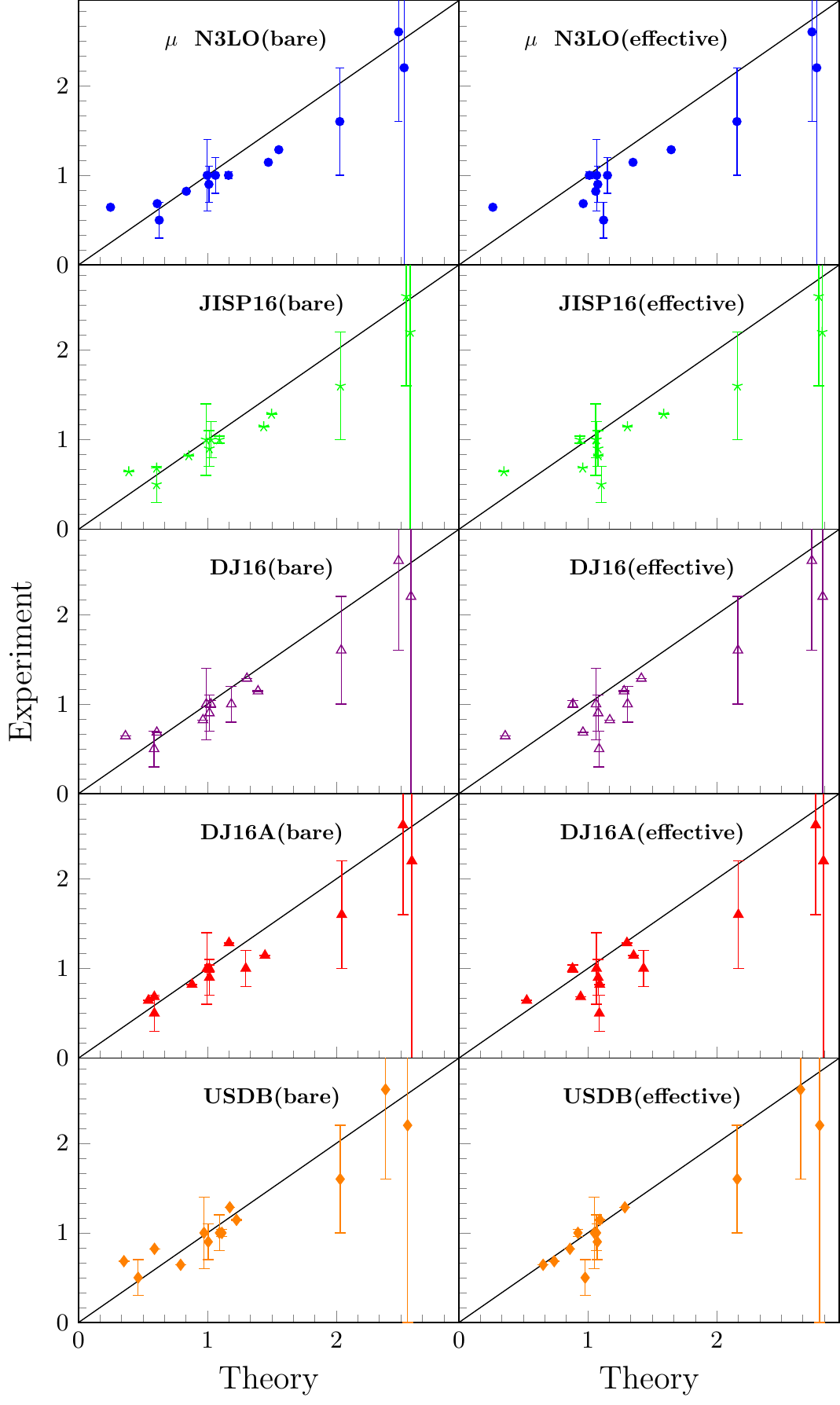}
		\caption{Comparison of experimental and theoretical magnetic moments with bare and effective values for the N3LO, JISP16, DJ16, DJ16A, and USDB interactions.}
		\label{Fig7}
	\end{center}
\end{figure*}

In Table \ref{tab2}-\ref{tab3}, computed shell-model quadrupole and magnetic moments are reported, where the experimental data of either quadrupole or magnetic moments or both are available. In Ref. \cite{Speidal_2008}, it is mentioned that proton excitation from $sd$- to $pf$-shell is less important in S and Ar than in the Ca isotopes. Thus, we want to test whether the interactions built for $sd$-shell are sufficient to describe the spectroscopy of those isotopes. First, we look at the quadrupole moments. We have given shell-model predictions for experimentally unavailable quadrupole moments of $^{32,36}$S and $^{38}$Ar. The experimental quadrupole moment for the first excited state of $^{32}$S is $-$0.16(2) eb. The  DJ16, DJ16A, and USDB interactions reproduce the correct sign, while calculations with the N3LO and JISP16 predict the opposite sign compared to the experiment for the $2^+$ state. In Ref. \cite{Saxena_Praveen_2017}, quadrupole moments for $^{32}$S and $^{33}$S have been calculated using IMSRG and CCEI approaches and concluded that IMSRG results are far from the experiment due to their wave functions. The respective quadrupole moments for $2^+$ state in $^{32}$S are $-$0.014 and $-$0.080 eb. In $^{34}$S, DJ16 and DJ16A fail to reproduce the sign of the quadrupole moment of the $2_1^{+}$ state. The shell-model predicts prolate structures of $^{35,36,37}$Cl isotopes, which also agrees with the available experimental one. For Ar isotopes, calculated quadrupole moments are slightly higher than the measured ones. The quadrupole moments are tabulated in Table \ref{tab2}. The magnitude of quadrupole moments obtained with optimized effective charges is slightly decreased compared to the moments with {standard} charges. The results obtained from different interactions are compared with the experimental data in Fig. \ref{Fig6}.

Now, we look at the magnetic moment results obtained from the shell-model calculations with bare g-factors. Previously, the magnetic moments for $^{32}$S are evaluated with Hamiltonians derived from IMSRG and CCEI approaches and are reported in Ref. \cite{Saxena_Praveen_2017}. Experimentally, magnetic moment for g.s. in $^{35}$S has been measured as 1.00(4) $\mu_N$. Our calculated results are closer to the experimental value. Earlier, the reported value with PSDPF interaction \cite{Bouhelal_2011} was 0.95 $\mu_N$ \cite{Grocutt_2022}. We have taken the experimental value of magnetic moments of first $2^+$ in $^{36}$S from Ref. \cite{Speidal_2008}, in which it was measured for the first time. In Ref. \cite{Speidal_2008}, the shell-model calculations with WBT interaction, two valence protons above the closed $d_{5/2}$ subshell, have been performed and it was obtained that the major configuration of this state was $\ket{\pi (s_{1/2}^1 d_{3/2}^1)}$ with the probability of 96.8\%. The magnetic moment of this state obtained with free g-factor was closer to the experimental one. Further, more extensive model space calculations, even excitations to the $f_{7/2}$ orbital, have been carried out, and they found that neutron excitations (to the $pf$-shell) are negligeble. Our obtained results with all microscopic interactions give the same configuration with 88-92\% probability and support the experimental value, and hence, the $N=20$ magic number. For $^{36}$S and $^{38}$Ar (both having $N=20$ closed neutron-shell), we see a significant change in the magnetic moment due to the two extra protons. The magnetic moments calculated with the $sd$ shell-model are in reasonable agreement with the measured one. In the case of sulphur and argon isotopes, a sudden change in the magnetic moment can be seen with the increase in neutron number from $N=18$ to $=20$, although it is slightly smaller in argon isotopes than sulphur. The reason is that for $^{36}$S, one proton is in $s_{1/2}$ orbital, and the other one is in $d_{3/2}$ orbital above $d_{5/2}^6$ coupled to $J=2$ and for $^{38}$Ar, the configuration is $d_{3/2}^2$, which makes a smaller magnetic moment than the previous configuration. In Ref. \cite{Speidel_2006,Speidal_2008}, $^{38}$S and $^{40}$Ar ($N=22$ nuclei) have been investigated and a very small g-factor is observed, which is due to the contribution from $fp$-shell neutrons in the wave functions. The g-factor for $^{32}$S and $^{36}$Ar ($N=Z$ nuclei) is measured around 0.5, which is confirmed by our results. Overall we find reasonable agreement with the experimental data. When we use effective g-factors as g$_{l,s}^{\rm eff}$ = (1.159, 5.15) for proton and g$_{l,s}^{\rm eff}$ = (-0.09, -3.55) for neutron \cite{Richter_2008}, then the magnetic moments deviate from the experimental values. A comparison between the experiment and theory is shown in Fig. \ref{Fig7} for magnetic moments. All interactions predict magnetic moments quite well with the experimental moments using bare g-factors.
\begin{table*}
	\setlength\tabcolsep{9.6pt}
	\centering
	\caption{\label{tab4}The rms deviations for different observables for {standard/optimized} effective charges and bare/effective g-factors corresponding to different interactions.}
	\begin{tabular}{lccccc}
		\hline \hline
		&	N3LO	&	JISP16	&	DJ16	&	DJ16A	&	USDB	 \T\B\\
		\hline 
		M(E2)(standard)	&	4.38	&	4.32	& 3.92	&	3.74	&	3.02 \T\\
		M(E2)(optimized)	&	3.93	&	4.13	& 3.37	&	3.12	& 3.10 \\
		Q({standard})	&	0.106	&	0.093	&0.078	&	0.061	&	0.057 \\
		Q(optimized)	&	0.102	&	0.090	&0.075	&	0.057	&	0.057 \\
		$\mu$(bare)	&	0.232	&	0.208	& 0.209	&	0.212	&	0.214 \\
		$\mu$(effective)	&	0.346	&	0.338	& 0.343	&	0.343	&	0.270 \B\\		
		\hline \hline		
	\end{tabular}
\end{table*}

In Ref.\cite{Speidel_2006}, it is noted that the B(E2) and E($2^+$) drastically change when one reaches the $N=20$ shell closure, but the magnetic moment does not, which indicates that these two different quantities have different sensitivity to the nuclear structure. The B(E2) value tells the combined collectivity of the nucleus, while the magnetic moment depends on specific proton and neutron configurations in the wave functions. 

We can say that the data of $2_1^+$ state in $^{36}$S and $^{38}$Ar can be reproduced without the excitation of $sd$-shell protons/neutrons to $pf$-shell, which support the $N=20$ magic number. For negative parity states, it is obvious that one needs to include $pf$-shell. Also, for excited states above the $2_1^+$ such as yrast $6^+$ in $^{36}$S, the inclusion of $pf$ shell is necessary. Beyond $N=20$ neutrons for S and Ar, the $sdpf$ calculations are important. In Ref. \cite{Kaneko_2011}, neutron-rich S and Ar isotopes were studied using the shell-model with a new effective interaction for $sdpf$-valence space, which was named as the extended pairing plus quadrupole-quadrupole forces accompanied by the monopole (EPQQM) interaction. The shell-model calculations reproduced the energy levels of these isotopes quite well, and B(E2) and quadrupole moments are also discussed. Neutron-rich sulphur isotopes, $^{38,40,42,44}$S, have been examined in Ref. \cite{Longfellow_2021} using SDPF-MU interaction. The shell-model calculations of magnetic moments for heavier even-even Ar isotopes ($^{38-46}$Ar) are reported in Ref. \cite{Robinson_2009} based on WBT \cite{Warburton_1992}, and SDPF-U \cite{Nowacki_2009} interactions and the role of configuration mixing is determined. In \cite{Bahr_2021}, near drip-line  $^{46-54}$Ar isotopes have been investigated in the framework of the shell-model using SDPF-U \cite{Nowacki_2009}, and SDPF-NR \cite{Nummela_2001} interactions. \\

{\bf The rms deviations:} For reduced E2 matrix elements, quadrupole moment, and magnetic moment, we have computed the rms deviations using the data given in their respective Tables \ref{tab1}-\ref{tab3}. The rms deviations are summarized in Table \ref{tab4} for standard/optimized effective charges and bare/effective g-factors. We have used central values to get rms deviations. There is a significant improvement for all microscopic interactions in the rms deviation for M(E2) when optimized effective charges are used. M(E2) values are highly sensitive to the used interaction. For N3LO, JISP16, DJ16, and DJ16A interactions with {standard} charges, the rms deviations are obtained as 4.38, 4.32, 3.92, and 3.74 $e$fm$^2$, respectively. The optimized effective charges reproduce the data with rms deviations of 3.93, 4.13, 3.37, and 3.12 $e$fm$^2$ for N3LO, JISP16, DJ16, and DJ16A interactions, respectively. The experimental quadrupole moment is not known for some nuclei, thus we have not included that data in the fit. Optimized effective charges also improve the rms deviation in the case of quadrupole moment. But, when we look at the rms deviation for the magnetic moment using effective g-factors, the situation is completely different. The deviation between experimental and theoretical moments for the effective g-factors increases, which means that the effective g-factors for microscopic interactions could be different from the USDB effective g-factors, taken from Ref. \cite{Richter_2008}. There is large error in experimental data for $4^+$ in $^{32}$S and $2^+_2$ in $^{38}$Ar. If we exclude these two data from the fit, an improvement in the rms deviations for DJ16A and USDB interactions is obtained as 0.149 and 0.162 $\mu_N$. By looking the Table \ref{tab4}, one cannot say which interaction is better among these interactions. 

\section{\label{IV}Conclusion}
We have performed shell-model calculations for sulphur, chlorine, and argon isotopes with $N\geq Z$ based on the microscopic effective $sd$-shell interactions, which are developed using the NCSM  and the OLS unitary transformation method. In the present calculations, we have used the N3LO, JISP16, DJ16, and DJ16A interactions and compared corresponding results with those obtained from the phenomenological USDB one. Based on the energies of excited states, we conclude that the DJ16A interaction provides the best results over other microscopic interactions. In addition to monopole modifications, more tuning in TBMEs of DJ16A interaction is required to reproduce the exact experimental results. Further, to check the reliability of these microscopic interactions, electromagnetic transition strengths, for which experimental data are available, are studied. The shell closure at $N=20$ is clearly visible in E($2^+$) and E2 transition strength for S isotopes. Some predictions for quadrupole moments based on these interactions are also made. Evaluated magnetic moments agree well with the experimental moments. Testing of these interactions on heavier $sd$-shell nuclei will be beneficial to determine further modifications in TBMEs. 

\section*{Acknowledgments}
We acknowledge a research grant from SERB (India), \\  CRG/2019/000556. P.C. acknowledges the financial support from the MHRD (Government of India) for her Ph.D. thesis work. We would like to thank N. A. Smirnova for her valuable suggestions. We acknowledge National Supercomputing Mission (NSM) for providing computing resources of ‘PARAM Ganga’ at Indian Institute of Technology Roorkee, which is implemented by C-DAC and supported by the Ministry of Electronics and Information technology (MeitY) and Department of Science and Technology (DST), Government of India.

\input{bibliography}

\end{document}

%% file: bibliography.tex
\bibliography{utphys}
\bibliography{references}